\documentclass{aa501}
\usepackage{graphicx}
\begin{document}
\title{Reliability of astrophysical jet simulations in 2D}
\subtitle{On inter-code reliability and numerical convergence}
\author{M.Krause
  \and{M.Camenzind}}
\institute{Landessternwarte K\"onigstuhl, D-69117 Heidelberg, 
  Germany}
\offprints{M.Krause,
M.Krause@lsw.uni-heidelberg.de}
\date{Received \today / Accepted <date>}
\abstract{%
In the present paper, we examine the convergence behavior and inter-code 
reliability of astrophysical jet simulations in axial symmetry. 
We consider both pure hydrodynamic 
jets and jets with a dynamically significant magnetic field. The setups were 
chosen to match the setups of two other publications, and recomputed with
the MHD code NIRVANA. We show that NIRVANA
and the two other codes give comparable, but not identical 
results. We explain the differences by the different application of 
artificial viscosity in the three codes and numerical details, which can be 
summarized in a resolution effect, in the case without magnetic field: 
NIRVANA turns out to be a fair 
code of medium efficiency. It needs approximately twice the resolution as the 
code by Lind
(Lind et al. 1989) and half the resolution as the code by K\"ossl
(K\"ossl \& M\"uller 1988). 
We find that some global properties of a hydrodynamical jet simulation, 
like e.g. the bow shock velocity, converge
at 100 points per beam radius (ppb) with NIRVANA. 
The situation is quite different after 
switching on the toroidal magnetic field: In this case, 
global properties converge even at 10 ppb. In both cases, details of 
the inner jet structure and especially the terminal shock region are still 
insufficiently resolved, even at our highest resolution of 70 ppb in the 
magnetized case and 400 ppb for the pure 
hydrodynamic jet. 
The magnetized jet even suffers from a fatal retreat of the Mach disk
towards the inflow boundary, which indicates that this simulation does not
converge, in the end.
This is also in definite disagreement with earlier simulations, and challenges
further studies of the problem with other codes.
In the case of our highest resolution simulation,
we can report two new features:  
First, small scale Kelvin-Helmholtz instabilities 
are excited at the contact discontinuity
next to the jet head.
This slows down the development of the 
long wavelength Kelvin-Helmholtz 
instability and its turbulent cascade to smaller wavelengths. 
Second, the jet head develops Rayleigh-Taylor instabilities which 
manage to entrain an increasing amount of mass from the ambient medium with 
resolution. This region extends in our highest resolution simulation over 2 
jet radii in the axial direction.   
}
\authorrunning{M. Krause \& M. Camenzind}
\titlerunning{Reliability of jet simulations in 2D}
\maketitle

\section{Introduction}

Since the publication of the ``twin exhaust model''
(\cite{BlandfordRees1974}), astrophysical jets 
have been  modeled by many workers.
These jets consist of a highly collimated outflow of magnetized plasma from 
a compact object, 
which -- in the case of extragalactic jets -- is assumed 
to be a black hole, and its accretion disk. 
In order to study 
the asymptotic propagation of 
such a plasma flow one needs a code 
that solves the equations of (magneto-) hydrodynamics
(MHD/HD) for the relevant initial 
and boundary conditions. Pioneers in this field were
M.\,L.\,Norman and coworkers in 1982 (\cite{Norman1982}). They were able to 
show that a flow of supersonic plasma remains stable and develops features 
that could be identified with features in the observations of radio galaxies:
A circular or conical deceleration area called {\it Mach disk} as the
hot spot, a strongly collimated beam as the elongated structure of the jets,
and a big zone of exhaust material as the lobes of the radio galaxies
(see also \cite{F98} for a recent review).
Meanwhile, more physics has been included in the calculations such as 
toroidal (perpendicular to the jet flow direction and the jet radius)
magnetic fields (\cite{Clarke1986}; Lind et al. 1989 (\cite{Lind1989})), 
poloidal magnetic fields
(\cite{Koessl1990}; \cite{Ryu1998}), special relativistic effects 
(\cite{Komissarov1999}; \cite{Aloy1999}), and the third dimension 
(\cite{Aloy1999}). For nonrelativistic hydrodynamical jets, the parameter 
space constituted by the Mach number and the ratio of the density of the 
jet to the external medium was explored up to high Mach numbers and low 
density ratios (\cite{M96}).  
  
Besides the numerical study of jets propagating in an undisturbed ambient 
medium, there has been considerable and increasing interest in the stability 
properties of jets during the last decade. The main agent of instability 
and possible destroyer of the jet is the Kelvin-Helmholtz (KH) 
shear instability.
A fundamental result of its linear analysis (\cite{AC92}; \cite{A96}) is that 
hydrodynamical jets without magnetic field are unstable to KH instabilities 
of a wide range of wavelengths, while jets with a poloidal field and even 
more those with a toroidal field in the cocoon (a distribution 
which is supported by simulation results, see \cite{Koessl1990})
are essentially stable to small wavelength perturbations.
Stability increases also with Mach number.
The development of long wavelength instabilities into the nonlinear regime 
was investigated
for the hydrodynamical case in cylindrical and slab symmetry and in
three dimensions by Bodo and coworkers (\cite{B95}, \cite{B94}, and \cite{B98},
respectively). They find that the instabilities destroy the jet 
in a time comparatively small with respect to the typical lifetime of 
an astrophysical jet source. This disruption could be proven to be less 
severe in the case where the jet is denser than the surrounding
medium, when radiative losses are taken into account (\cite{Mic00}, for the 
three dimensional case), and is even impeded if one 
includes an equipartition magnetic 
field (\cite{HCR97}, three dimensional, poloidal fields; 
\cite{RHCJ99}, three dimensional, also toroidal fields).

With the exception of the latter authors, all of the above mentioned 
simulations were conducted using only one resolution level.
This resolution level seems to be quite arbitrary and
is not upgraded with the years. For example, while K\"ossl and M\"uller 
(1988) (\cite{Koessl1988}) 
considered a resolution of 100 ppb, 
both in radial and longitudinal direction as insufficient
to resolve the dynamical structures of their hydrodynamical jet, 
Massaglia et al. (1996) considered their scaled grid with 20 points
in the radial and 11 points in the longitudinal direction 
at maximum of their likewise hydrodynamical simulation as ``high resolution'',
which should be -- due to their superior code -- about the same.
This is due to the above mentioned increase of physical ingredients
into the simulation. To give another example: \cite{Lind1989}
and Rosen et al (1999) use with a comparable numerical scheme for a
magnetized jet in two and three dimensions respectively both 15 ppb
to resolve the transversal direction.

A reliable numerical simulation of an astrophysical jet has to be converged 
regarding its internal structure as well as the behavior of its boundary.
This is true also for the study of surface instabilities because
the jet body behavior can influence its surface. 
Furthermore, in the literature 
it is normally assumed that long wavelength modes can be studied independently
from shorter wavelength modes. The validity of the latter assumption is 
particularly questionable in the hydrodynamical case, and not so much for 
the magnetized case as linear stability analysis shows, as mentioned above,  
that small wavelength perturbations to the surface 
are stabilized in a magnetized jet. This should be reflected in the resolution
that is needed in a simulation in order to catch the relevant physics,
especially in a situation, where small scale and large scale behavior 
could influence one another.

One aim of the present paper is therefore the investigation of the convergence 
behavior  and the role of 
small scale structure of both the hydrodynamical (Sect. 3.4) 
and the magnetized case (Sect. 4.3).
The computations are carried out with the MHD code 
NIRVANA and are compared with 
simulations from the literature.
Since in such an investigation high resolution is essential, we restrict 
ourselves to the two dimensional axisymmetric case. This is justified 
by the fact that three dimensional simulations show more instability
but do not differ essentially from the two dimensional ones.
Even with this restriction we needed 3 months of CPU time on a Pentium III
workstation to perform our highest resolution model with 
6.4 million grid points. 

A hydrodynamic or MHD code 
constructed after the van Leer scheme (e.g. the famous ZEUS code)
is
certainly less effective than a code with a piecewise parabolic method 
(\cite{Koessl1988}; \cite{WC84}).
But up to now, there is no comparison of the results of  different 
van Leer scheme codes available 
for astrophysical jet simulations. 
However, Woodward and Colella (1984) showed that besides strong differences 
in a 1D test problem, the second order accurate codes they tested performed overall
equally well in the 2D case, although differences occurred in some details. They note
(\cite{WC84}, p166): 
{\em Does the accurate representation of a jet in one part of the flow
  compensate for the presence of noise in another part?}
Depending on the problem, this really could make a difference.
In this paper, we compare the results of different MHD codes for 
the special case of astrophysical jet simulations, 
in Sect. 3.2 and 3.3 for the 
HD, and in Sect. 4.2 for the MHD case. For this purpose,
we recompute the results from two previous publications with the 
MHD code NIRVANA (\cite{Ziegler1997}), and analyze 
the differences to the published, original, results.
\begin{figure*}
\centering
 \includegraphics[width=17cm]{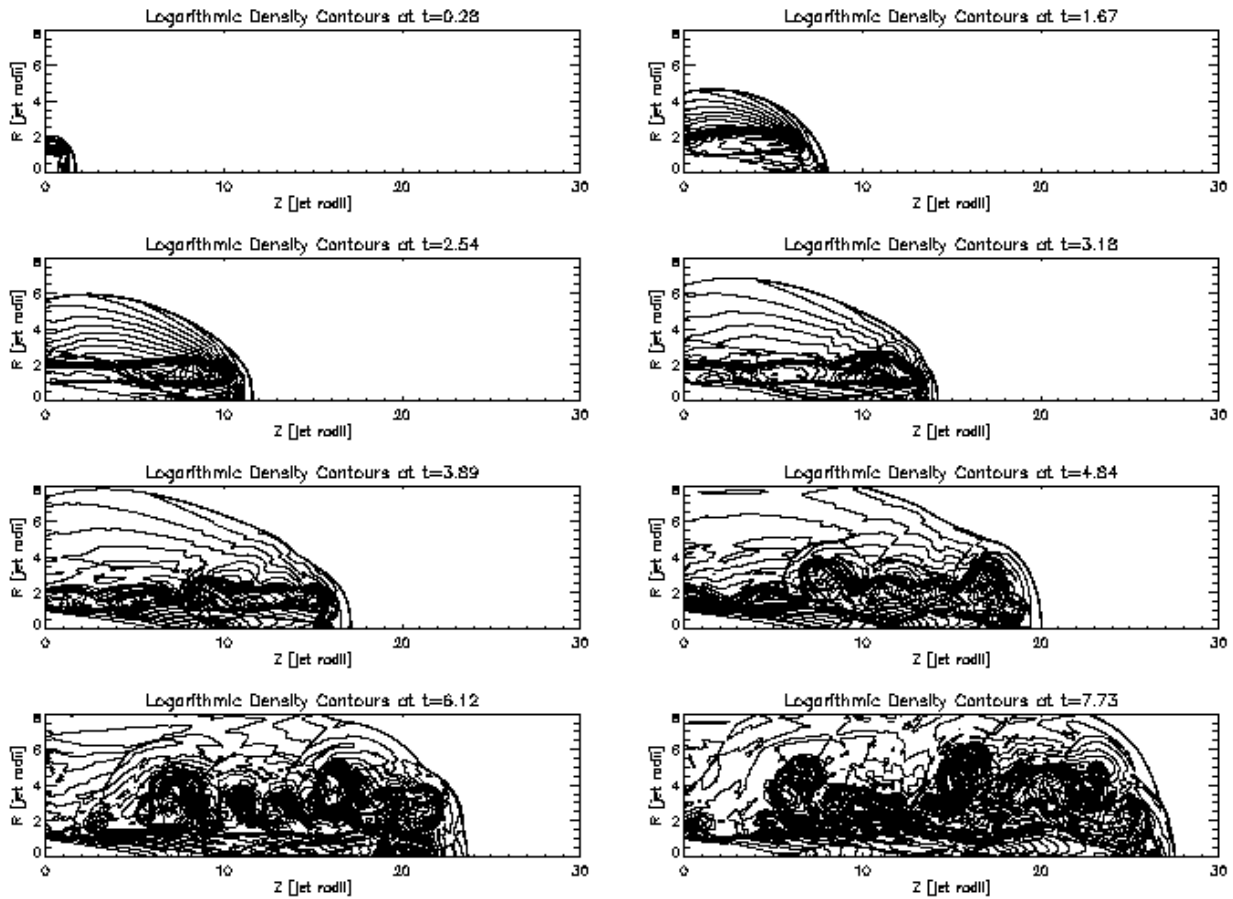}
 \caption{Contour plots of the density (30 logarithmically spaced lines)
    for the jet model of setup A. The times for the snapshots were chosen
    in order to match Fig. 11 of \cite{Koessl1988} closely.}
  \label{kmden}
\end{figure*}
  
\section{Description of the codes}
The reference codes are a two dimensional MHD code by Lind
(\cite{Lind1986}; \cite{Lind1989}), named {\it FLOW}, 
and an also two dimensional
HD code by K\"ossl (\cite{Koessl1988}), hereafter {\it DKC}.
A precise description of FLOW can be found only in Lind's Ph.D. thesis
(\cite{Lind1986}). There, Lind points out two special features of FLOW.
First, it uses a predictor corrector algorithm for the timestep 
calculation and second, a specific advection method is applied.
This method considers the matter density flux as the primary advective
quantity and calculates the remaining fluxes by multiplying the matter 
density by the specific density in the cell origin. It is not a
Godunov scheme, and does not use an exact or approximate Riemann solver.
The code that we used for 
the recomputations was {\it NIRVANA} (\cite{Ziegler1997}),
capable of three dimensional computations but used here in the 2D mode.
The latter has like the others accomplished the usual tests and has already
been used in simulations of proto-stellar jets (\cite{Thiele2000})
and other astrophysical problems (e.g. \cite{ZU97}).
All these codes use explicit Eulerian time stepping.
They are second order accurate and use a monotonic upwind differencing scheme.
They treat 
the following standard set of ideal (magneto-) hydrodynamic equations:
\begin{eqnarray}
\frac{\partial \rho}{\partial t} + \nabla \cdot \left( \rho {\bf v}\right)&
 = & 0 \\
\frac{\partial \rho {\bf v}}{\partial t} + \nabla \cdot\left( \rho {\bf v} 
{\bf v} \right) & = &
- \nabla p + \frac{1}{4 \pi} \left({\bf B} \cdot \nabla \right) {\bf B}
-\frac{1}{8 \pi} \nabla {\bf B}^2 \\
\frac{\partial e}{\partial t} + \nabla \cdot \left(e {\bf v} \right) & = &
- p \; \nabla \cdot {\bf v} \label{ie}\\
\frac{\partial {\bf B}}{\partial t} & = &
\nabla \times ( {\bf v} \times {\bf B}), 
\end{eqnarray}
where $\rho$ denotes the density, $e$ internal energy density, 
${\bf v}$ velocity,
${\bf B}$ the magnetic field and $p=(\gamma -1) e$ the pressure. Here
$\gamma = 5/3$ for a nonrelativistic monoatomic gas is assumed. Instead of the 
internal energy $e$, FLOW and DKC use the total energy 
$u=\rho v^2/2+e+B^2/8\pi$. Equation (\ref{ie}) is then replaced by:
\begin{equation}
\frac{\partial u}{\partial t} + \nabla \cdot \left(u {\bf v} \right) =
- \nabla \cdot (p {\bf v}).
\end{equation}
This is analytically equivalent, but may 
give different numerical results, in particular 
in regions of discontinuous flows. 
Another difference of the codes is the 
use of artificial viscosity. 
FLOW has no need for artificial viscosity at all, 
according to test calculations reported in Lind (1986).
DKC and NIRVANA use an artificial 
viscosity in order to enhance the diffusion in regions of strong gradients.
This has the effect that shocks are transported correctly without numerical
oscillations at the cost of smoothing the shocks 
over some grid zones. DKC even makes use of an antidiffusion term, which
cancels the effects of artificial viscosity in regions of smooth flow.
This point reflects the differences in the details 
of the implementation of the three codes as summarized in Table 1.
\begin{minipage}{\hsize}
\vspace{3mm}
{\bf Table 1.} Differences between the discussed codes\nopagebreak
\vspace{1mm}\nopagebreak \noindent
\begin{center} \noindent
\begin{tabular}{lccc}  \hline \hline 
\multicolumn{4}{c}{\vspace{-3mm}}\\
                 & FLOW & DKC & NIRVANA \\ \hline
  uses e         &      &     &    x    \\ 
  uses u         &   x  &  x  &         \\ 
  art. viscosity &      &  x  &    x    \\ 
  antidiffusion  &      &  x  &         \\ \hline \hline
\end{tabular}
\end{center}
\vspace{0mm}
\end{minipage}

\section{Hydrodynamic jet simulations} 
\begin{figure*}
\centering
 \includegraphics[width=17cm]{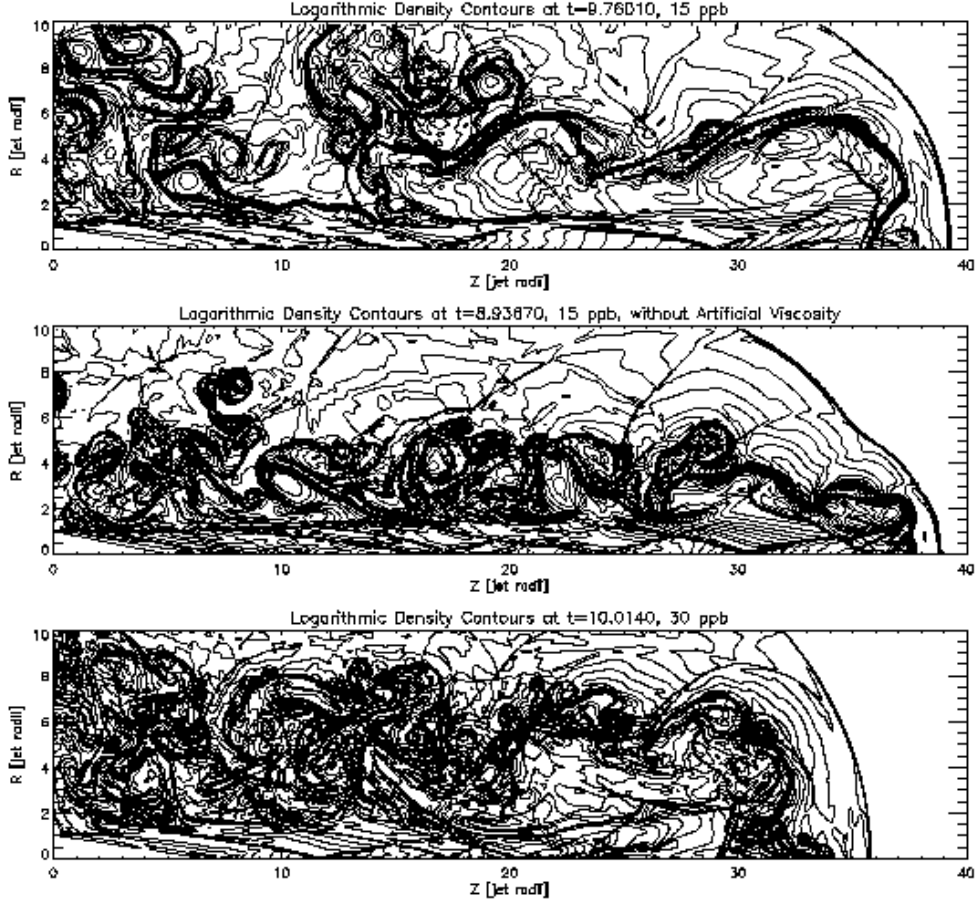}
  \caption{Logarithmic density plot of the jet from setup C. 
    Top: 15 ppb resolution, t=9.76; 
    middle: 15 ppb resolution without artificial viscosity, t=8.94; 
    bottom: 30 ppb resolution, t=10.00}
  \label{lhrc}
\end{figure*}
\subsection{Numerical Setup}
As is appropriate for the simulation of axially symmetric jets
we use (normalized) cylindrical coordinates (r,z).
The jet is injected into a homogeneous ambient medium. 
We take the density and the pressure there as the unit values.
Therefore the sound speed in this external medium is $\sqrt{\gamma}$.
The jets are injected in pressure equilibrium and have a density contrast
of $ \eta = 0.1$ 
which means that the jet material has 0.1 times the density of 
the ambient medium. The unit of length is the jet radius $R_\mathrm{j}$.
Velocity is measured in units of $\sqrt{p_\mathrm{m}/\rho_\mathrm{m}}$, 
where $p_\mathrm{m}$
is the pressure and $\rho_\mathrm{m}$ the density in the ambient medium,
and the time unit is $R_\mathrm{j} / \sqrt{p_\mathrm{m}/\rho_\mathrm{m}}$. 
The units were chosen to match the normalized units in \cite{Lind1989} and
\cite{Koessl1988} as closely as possible.
The boundary conditions are
rotational symmetry at the jet axis, an outflow condition at the upwind 
boundary for setup A and an impenetrable wall for 
setup B and C (except for the nozzle). 
Open boundaries were applied
on the two other sides besides setup C where outflow was specified 
in order to match the original conditions. This should have no noticeable
effect, since inflows from those sides are hardly expected.
The simulations are carried out at different resolution characterized by
the number of grid points the jet beam is resolved with (ppb).
We also add higher resolution images than the original computations.
Setup A and B recompute the results from \cite{Koessl1988},
where two simulations with higher resolution were added to setup B, and
setup C is designed according to the hydrodynamic jet from \cite{Lind1989}.
The parameters are summarized in Table 2.
We note the following on the jet nozzle: 
In our MHD simulations we encountered problems
when the boundary condition was applied in the usual one grid zone only.
Therefore we fixed the jet values in an area of four cells from the upwind 
boundary. In the simulations without magnetic field, we did this in the 
same way, although it did not influence our results here.
\begin{minipage}{\hsize}
\vspace{3mm}
{\bf Table 2.} Parameters of the jet models
\vspace{-3mm}
\begin{center}
\begin{tabular}{lccc}  \hline \hline
\multicolumn{4}{c}{\vspace{-3mm}}\\
  Setup                &   A     &    B         &     C   \\ \hline
  Jet velocity $v_\mathrm{j}$   &  24.5   &  24.5        & 25.0    \\
  Resolution (ppb)     &  40     &  10,20,40,70,& 15,30 \\
                       &         &  100,200,400 &     \\ 
  upwind boundary      & outflow &  reflecting  &  reflecting   \\
  comp. area (in $R_\mathrm{j}$)& 8$\times$30 & 4$\times$10 &
  10$\times$40\\ 
  reference code       & DKC     & DKC     & FLOW    \\ \hline \hline
\end{tabular}
\end{center}
\vspace{0mm}
\end{minipage}

\subsection{Setup A: inter-code comparison of a time series}
\begin{figure}
\resizebox{\hsize}{!}{\rotatebox{-90}{ \includegraphics{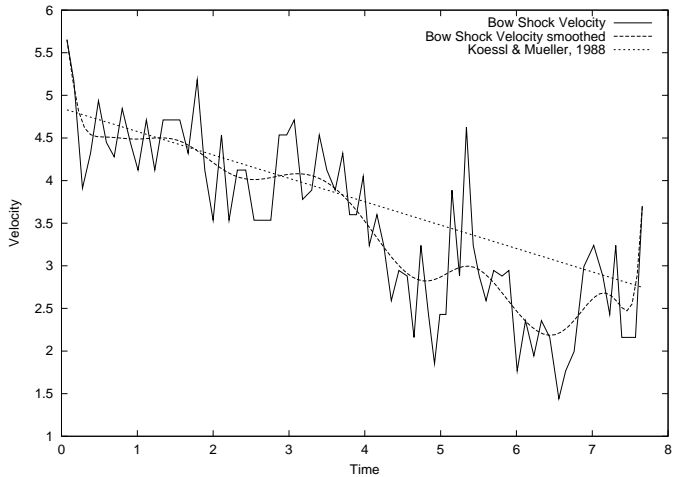}}}
 \caption{Bow shock velocity over time for the jet of setup A (solid line), 
    smoothed (long dashed line) and for comparison the approximate behavior
    in KM88 adopted from their figure 9a (short dashed 
    line). 76 time steps were used for this graph.}
\label{kmbow}
\end{figure}

The time evolution of the jet model of setup A is shown in Fig.~\ref{kmden}.
Snapshot times were chosen very similar to the corresponding model in 
\cite{Koessl1988}. Comparing the two computations, at the first glance,
one recognizes a great similarity: 
In both cases at early times, a laminar backflow
is established, where the terminal Mach disk is nearly perpendicular 
to the z-axis.
At $t=3.18$ the laminar backflow phase terminates with the Mach disk 
becoming oblique and one of the forward and backward moving shocks from 
the beam
extending into the backflow where it decelerates the backflow deflecting 
it away
from the axis. The material to the left of this shock structure at approximately
$z=8$ at $t=3.89$ is pumped off the grid and has disappeared at $t=6.13$.
The rest of the backflow material essentially stays at its place until the 
end of the 
simulation expanding from 2 up to 6 jet radii. Also visible in our simulation
are the crossing oblique shock waves in the beam. 
So far the described features
matched the corresponding ones in the original publication very well.
In Fig.~\ref{kmbow} the bow shock velocity from our computation is compared 
to the one in the original publication. 
It shows a variation on the level of every
evaluated timestep with an amplitude of roughly one in the Mach number. 
This was unrevealed in the original paper due to the lower 
sampling rate (23 versus 76 snapshots in our computation). The oscillation is 
modulated by a mode with longer wavelength (represented by the smoothed curve 
in Fig.~\ref{kmbow}) which is about 2 in our results and 1.7 in 
\cite{Koessl1988}. They called this periodic de- and reaccelleration 
``beam pumping''. It was interpreted by Massaglia et al. (1996) in the 
following manner: The oblique shocks have a high pressure on the axis. 
Each time they arrive at the jet head, the head is accelerated 
due to this pressure gradient.
Because of this phenomenon multiple bow shocks appear
which can be seen in Fig.~\ref{kmden} (especially at $t=6.13$).
The overall jet velocity decreases in the laminar phase in a 
similar way as in the DKC computation. But after $t \approx 4$ the decrease 
in our computation is faster. At $t=7.73$ our jet has approached $z=27.43$
which is $9\%$ less than the reference value.
At a first glance it is surprising that the jet velocity decreases at all,
for it can be derived very easily
by equating the ram pressure
of the jet, $\rho_\mathrm{j} (v_\mathrm{j} - v_\mathrm{bow})^2$ 
to the pressure in the external medium
$\rho_\mathrm{m} v_\mathrm{bow}^2$ 
that the bow shock velocity should be constant:
\begin{displaymath}
v_\mathrm{bow} \approx \frac{\sqrt{\eta}\, v_\mathrm{j}}{1+\sqrt{\eta}},
\end{displaymath}
where $\eta = \rho_\mathrm{j} / \rho_\mathrm{m}$ is the density contrast. 
For the jet under consideration this bow shock velocity should evaluate to
$5.9$, approximately achieved in very early times of our simulation.
The explanation for this decrease in the bow shock velocity was given 
by \cite{Lind1989}: As the jet propagates, its head 
expands and $\eta$ in the above formula has to be replaced by $\eta \epsilon$
with $\epsilon = A_\mathrm{j} / A_\mathrm{head}$ the ratio between 
the area of jet beam and its head. Taking this into account we conclude
that the effective working surface for the jet is $\approx 3 R_\mathrm{j}$ 
at late 
times of our simulation. This corresponds well to the extent of the 
jet head in our contour plots (Fig.~\ref{kmden}). 
A clear discrepancy between the simulations is the appearance of the 
KH instabilities. Where in the original publication only
an unstructured bump is visible, in our computation a round filigree system
with additional instabilities can be seen. At this point we propose that 
the observed differences are caused by a different effective resolution of 
the two codes (This question will be examined in more detail in Sect. 3.4). 
So one can say that the qualitative
behavior of the simulation is well reproduced, which gives additional 
confidence in the quality of both simulations. But on a quantitative level
there are differences. The NIRVANA jet is slower at late times and develops 
a richer cocoon structure. The differences arise in the non-laminar flow phase. 
\subsection{Setup C: the effect of artificial viscosity}
\begin{figure*}
 \centering
 \includegraphics[width=17cm]{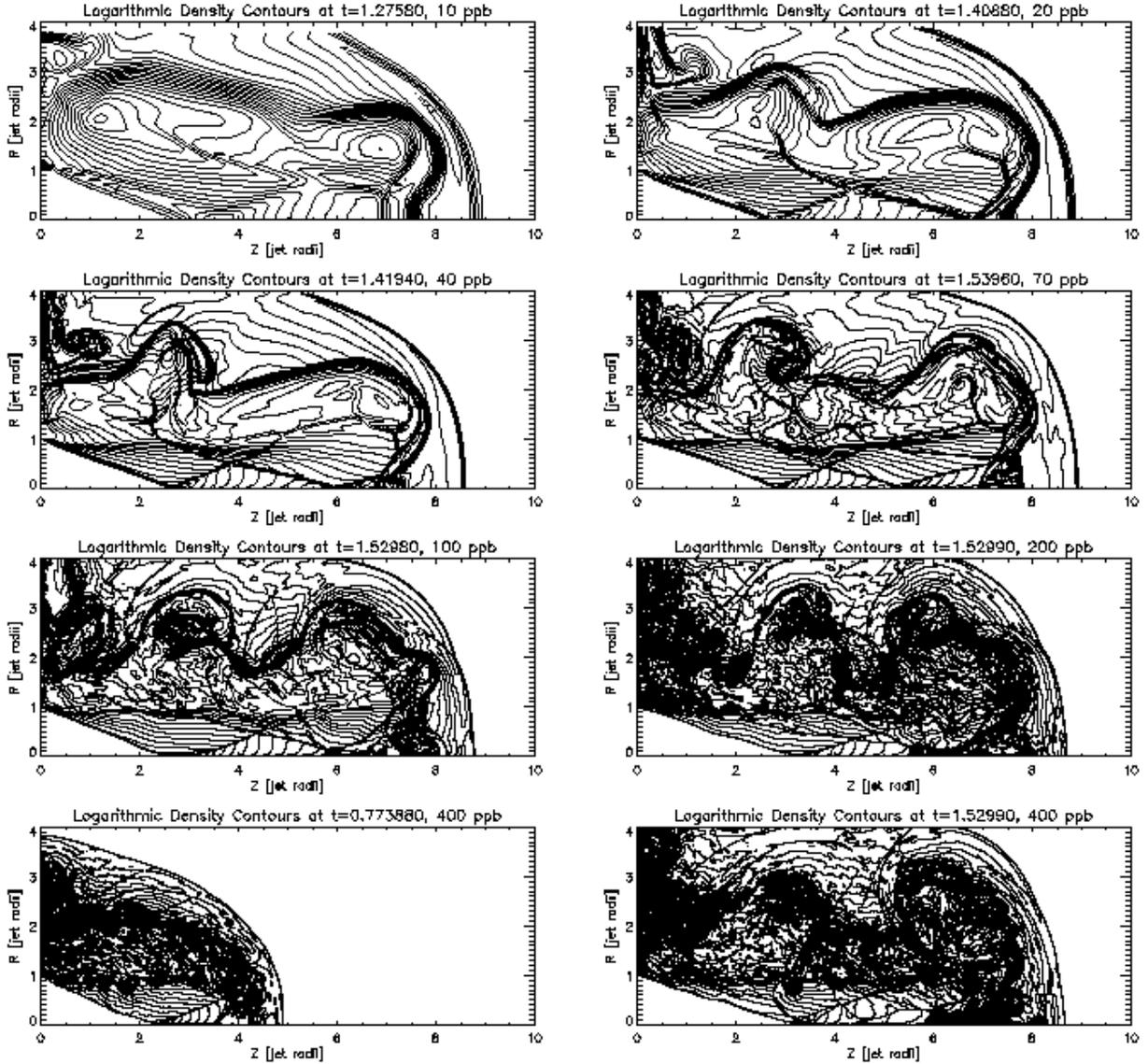}
  \caption{Jets from setup B. 30 logarithmically spaced density contours are 
    shown. Time and resolution are indicated on top of the individual
    pictures.}
  \label{kmrescomp}
\end{figure*}
\begin{figure}
 \resizebox{\hsize}{!}{\rotatebox{-90}{\includegraphics{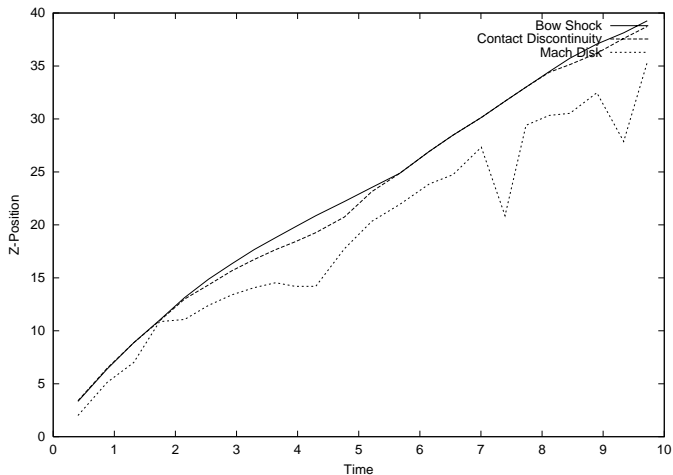}}}
 \caption{Flow parameters for the hydrodynamic jet model of setup C
   at 15 ppb with shock handling by artificial viscosity.
   The vertical axis displays the z-coordinate and the horizontal axis 
   the time in our units which is 1/15 of Lind's time unit. 
   The highest z-value has always the bow shock, the contact 
   discontinuity has a lower or equal z-value and the Mach disk takes
   the lowest value.}
  \label{lhbow15}
\end{figure}

The reference paper for this subsection is \cite{Lind1989}.
They plot their hydrodynamic jet model at t=10 (in our units).
This is not possible for our recomputation because at t=10 the NIRVANA jet
has left the computational domain. For that reason, we discuss the last
timestep with the bow shock not having 
left the grid ($t=9.76$, see Fig.~\ref{lhrc}).
At a first glance, it seems that our jet differs quite a lot from the 
reference jet: It propagates faster (average velocity of 4 versus 3.5), 
the region between bow shock and contact 
discontinuity on the axis (at z $\approx$ 36 in Fig.~\ref{lhrc}) extends 
over one jet radius (versus three, compare Fig.~\ref{lhbow15}),
the inner beam shock structure is by far more irregular in our computation
and this prevents the cocoon from developing regular vortices.
In the original publication the cocoon extended everywhere over about 
7 jet radii, whereas at the end of our simulation one vortex has even 
started to leave the grid over the upper boundary.
Why are there such strong differences? In order to investigate this question 
we performed the simulation again 
with twice the resolution (Fig.~\ref{lhrc}).
The result is striking. The average velocity reduces to 3.6 which is only
2\% higher than the 3.5 reference value. The region between bow shock and 
contact discontinuity is amplified to two jet radii 
(compare Fig.~\ref{lhbow30}),
the cocoon extends over 8 jet radii, approximately, and in the interior of the 
beam there are more oblique shocks. 
This strongly suggests that with a little bit more than twice the resolution
NIRVANA reproduces the global parameters achieved in the FLOW simulation. 
Another reason for the differences
might be the shock handling by methods of artificial viscosity in NIRVANA.
Therefore, we have repeated the simulation at the original resolution (15 ppb),
but without artificial viscosity. 
We note here that artificial viscosity is not an option but a necessity
in order to handle shocks correctly for NIRVANA. Fig.~\ref{lhrc} shows 
that there is now almost nothing of the cocoon 
material accumulating on the left hand boundary. Instead most of it 
is consumed in KH instabilities of approximately equal wavelength 
than in the original publication. There are more oblique shocks 
in the beam, too. We conclude,
that Lind's hydrodynamic jet simulation is, strictly speaking, not 
reproducible by our code. But we can explain the 
shape of the cocoon by the 
different diffusivity description in Lind's 
code and the global 
parameters by a doubling of the resolution in NIRVANA.
The comparison between our 15 ppb model and our 30 ppb model clearly
shows a strong dependence of the growth of the KH instability
on the resolution.

\subsection{Setup B: the numerical convergence of a hydrodynamic jet
 in comparison to \cite{Koessl1988}}
\subsubsection{Details of the flow}
The results are shown in a series of logarithmic density plots
in Fig.~\ref{kmrescomp}. The times for the snapshots were chosen in order to 
match the one in the original computation 
with an accuracy of better than one percent.
Comparing the contour plots with the results from \cite{Koessl1988}
one can immediately see that they are very similar with the exception that 
we needed only about half the resolution to achieve the same result:
At the resolution of 10 ppb we get the conspicuous ledge at 
$(r,z)\approx(1.5,8)$
of the contact discontinuity which appears only at 20 ppb in the original.
The hook like structure at $(r,z)\approx(3,1)$ can be seen for the first time 
in the 20 ppb plot at ours and looks almost identical to the 40 ppb picture 
in \cite{Koessl1988}. With NIRVANA we are able to resolve 
internal shock waves with our lowest resolution (10 ppb versus 20 ppb). 
Crossing shock waves 
become visible at 20 ppb (versus 40 ppb) and the upper right one of the 
cross like shocks appears at 40 ppb (versus $\approx$ 70 ppb).
The inner beam also consists of plane and centered rarefaction waves (see 
\cite{Koessl1988} for details) which one can identify at 10 ppb in our 
simulation (versus 20 ppb).
The cocoon is dominated by a prominent KH instability of 
wavelength $\approx 4$ in the original publication 
which becomes somewhat shorter 
in our simulations. It appears as a break in the contact discontinuity 
at low resolution. At 70 ppb in our simulation a round structure has formed
which, approximately, retains this size up to 200 ppb. Only a part of the 
interior of this structure develops higher mode KH 
instabilities. This means that this region becomes turbulent.
No sign of that can be seen in the DKC simulations. 
They resolve 
KH instabilities on the level of breakpoints in the contact 
discontinuity even at 100 ppb.
\begin{figure*}
\begin{minipage}{17cm}
\rotatebox{0}{\includegraphics[width=8.5cm]{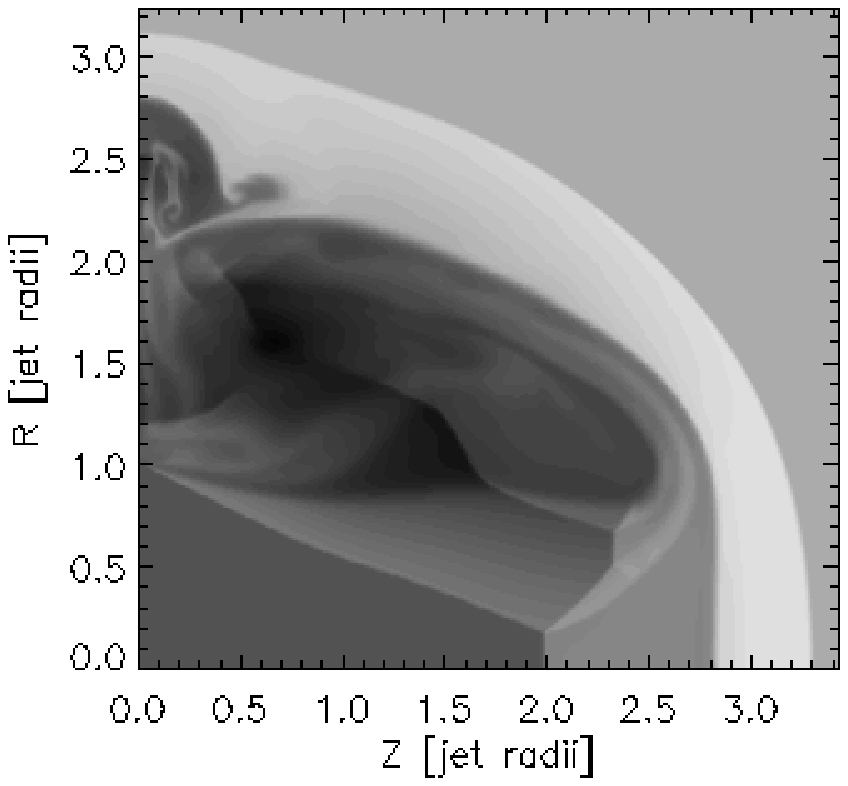}}%
\rotatebox{0}{\includegraphics[width=8.5cm]{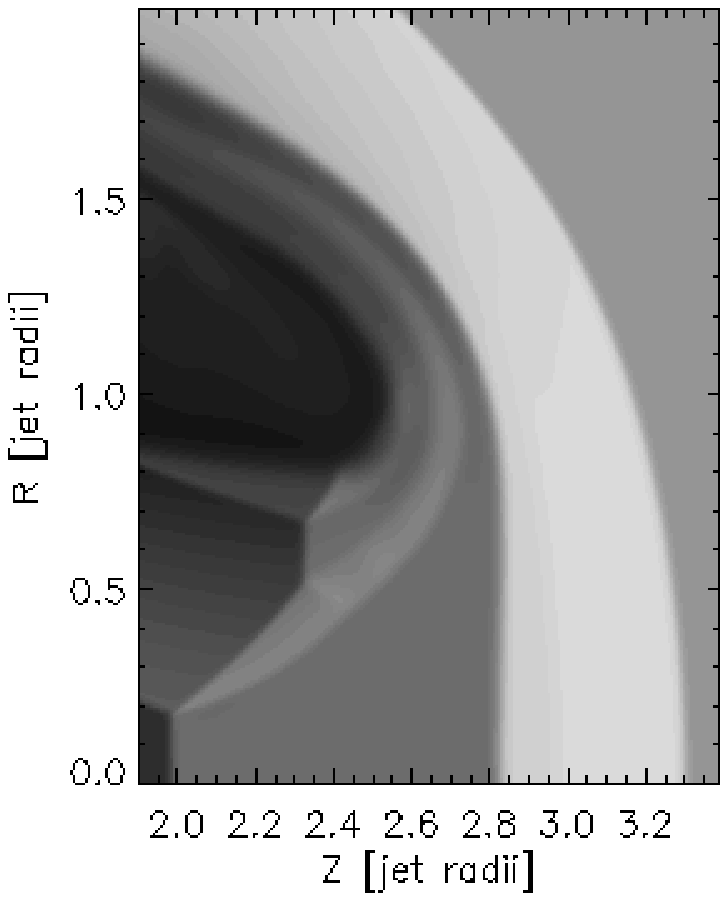}}\\[0.2cm]
\mbox{}\hfill a \hfill\hfill b \hfill\mbox{}\\[0.4cm]
\rotatebox{0}{\includegraphics[width=8.5cm]{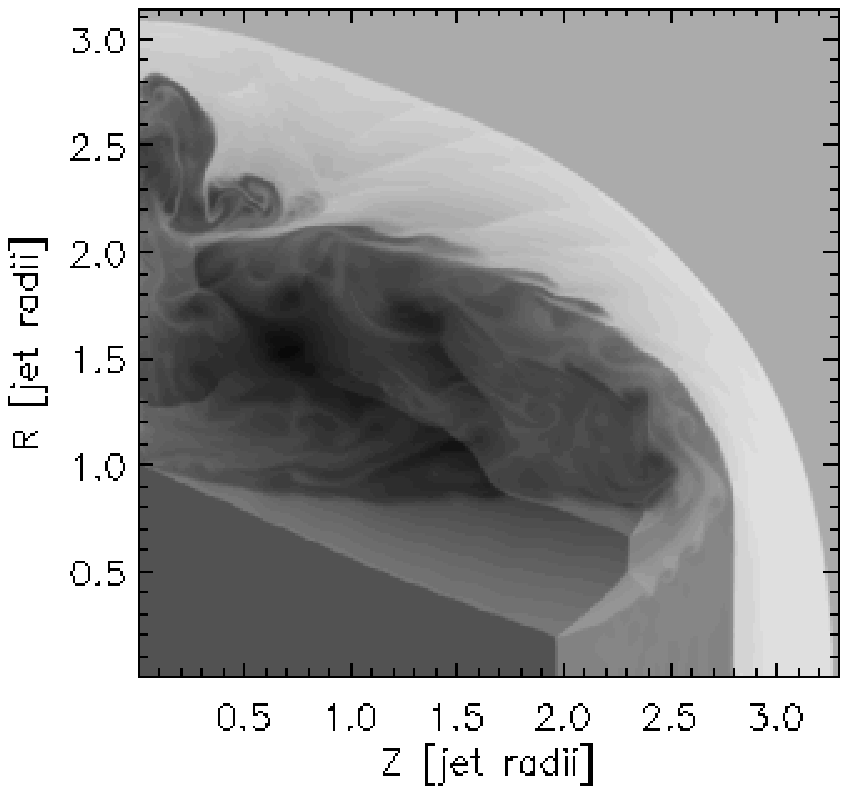}}%
\rotatebox{0}{\includegraphics[width=8.5cm]{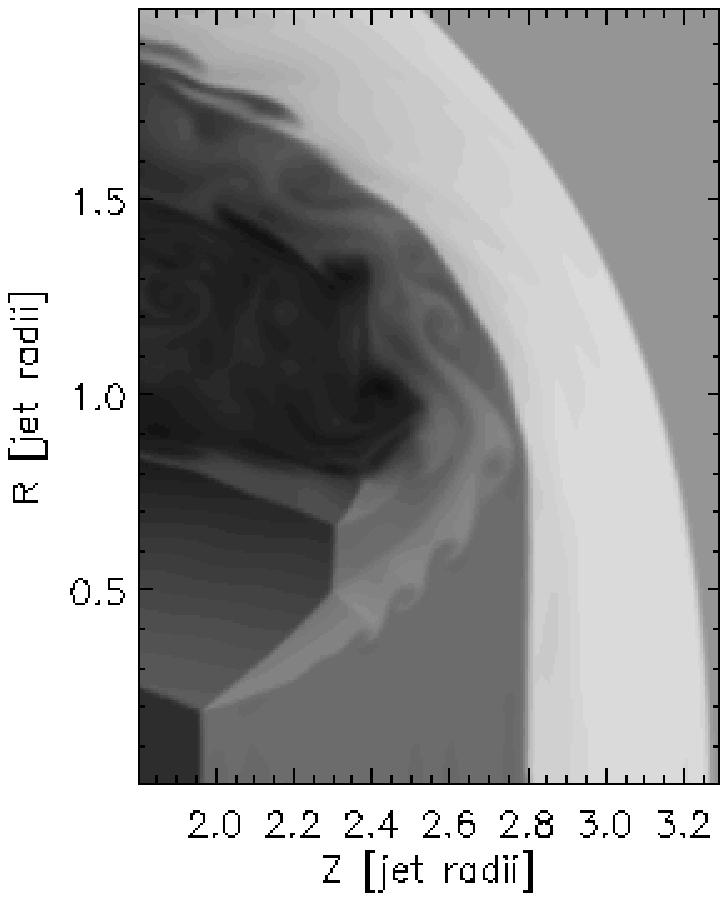}}\\[0.2cm]
\mbox{}\hfill c \hfill\hfill d \hfill\mbox{}\\[0.4cm]
\end{minipage}
 \caption{Gray scale plots of the logarithmic density for the jets of setup B.
   On the right hand side are magnifications of the head region of the 
   corresponding picture to the left. The upper (a and b) and lower (c and d)
   pictures show 
   the 100 ppb and 200 ppb simulations, respectively. The time is 0.500 in
   all cases.}
  \label{kmrcgray_a}
\end{figure*}
\begin{figure}
 \resizebox{\hsize}{!}{\rotatebox{-90}{\includegraphics{fig07.epsi}}}
 \caption{Same as Fig.~\ref{lhbow15}, but at a higher resolution of 30 ppb.}
  \label{lhbow30}
\end{figure}

\subsubsection{Kelvin-Helmholtz instabilities at the contact discontinuity
  and beam boundary}
\begin{figure*}
\begin{minipage}{17cm}
\rotatebox{0}{\includegraphics[width=8.5cm]{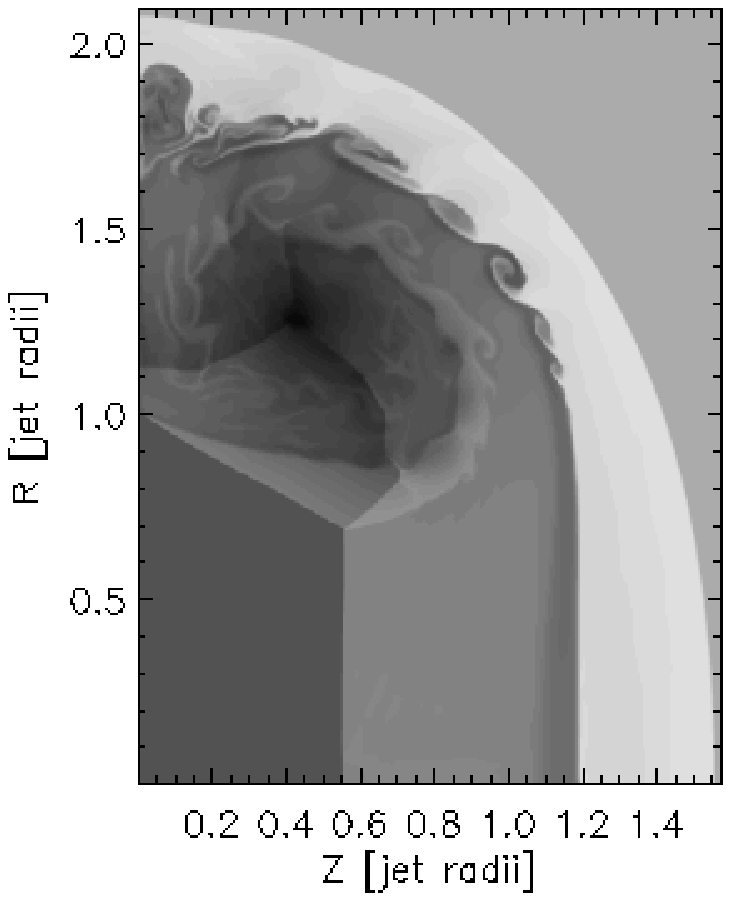}}%
\rotatebox{0}{\includegraphics[width=8.5cm]{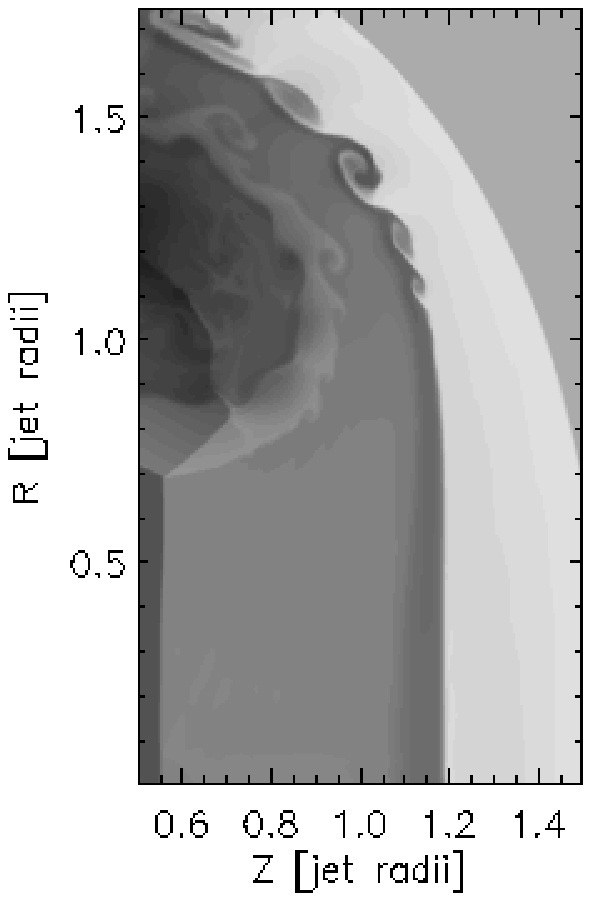}}\\[0.2cm]
\mbox{}\hfill a \hfill\hfill b \hfill\mbox{}\\[0.4cm]
\rotatebox{0}{\includegraphics[width=8.5cm]{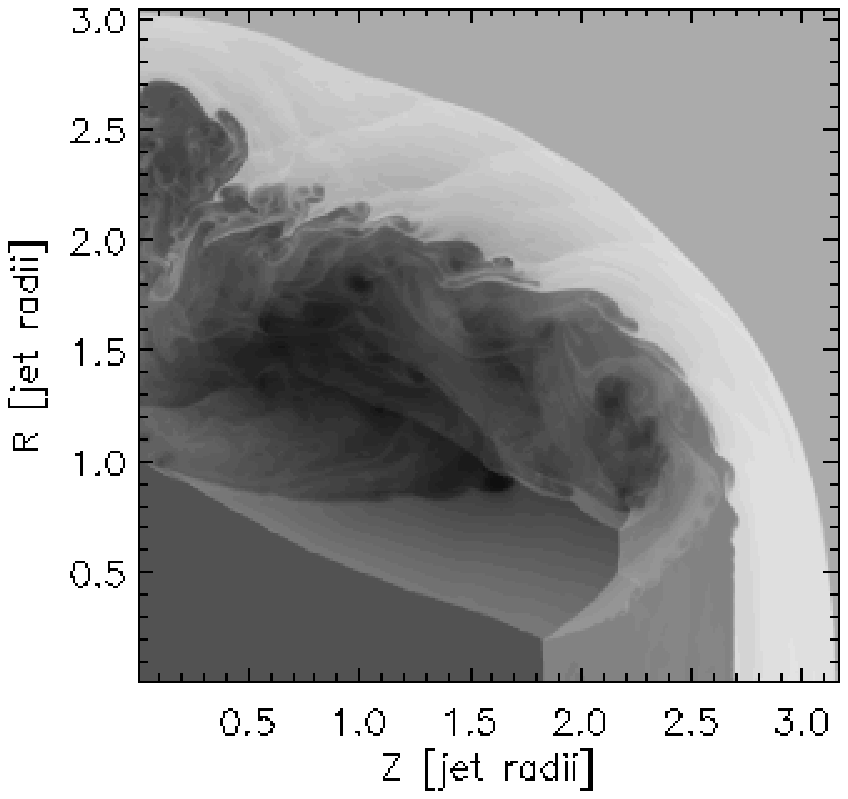}}%
\rotatebox{0}{\includegraphics[width=8.5cm]{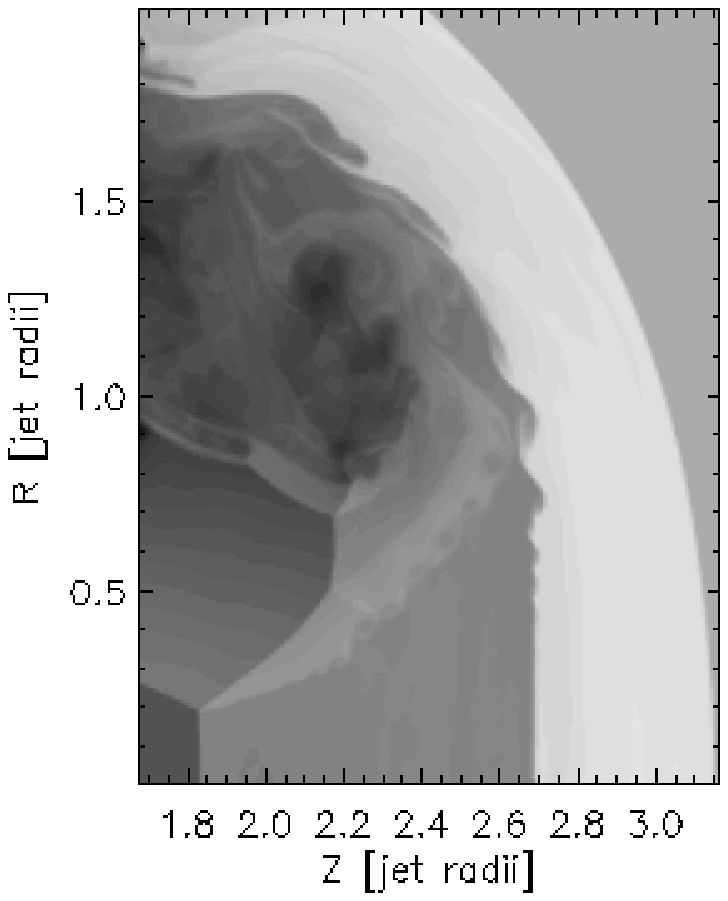}}\\[0.2cm]
\mbox{}\hfill c \hfill\hfill d \hfill\mbox{}\\[0.4cm]
\end{minipage}
 \caption{Same as \ref{kmrcgray_a} but for 400 ppb and at times t=0.216
   (upper pictures) and t=0.483 (lower pictures). (The data in this figure
   and in Fig.~\ref{kmrcgray_a} was rebinned to 100 ppb for visualization.)}
  \label{kmrcgray_b}
\end{figure*}

\begin{figure*}
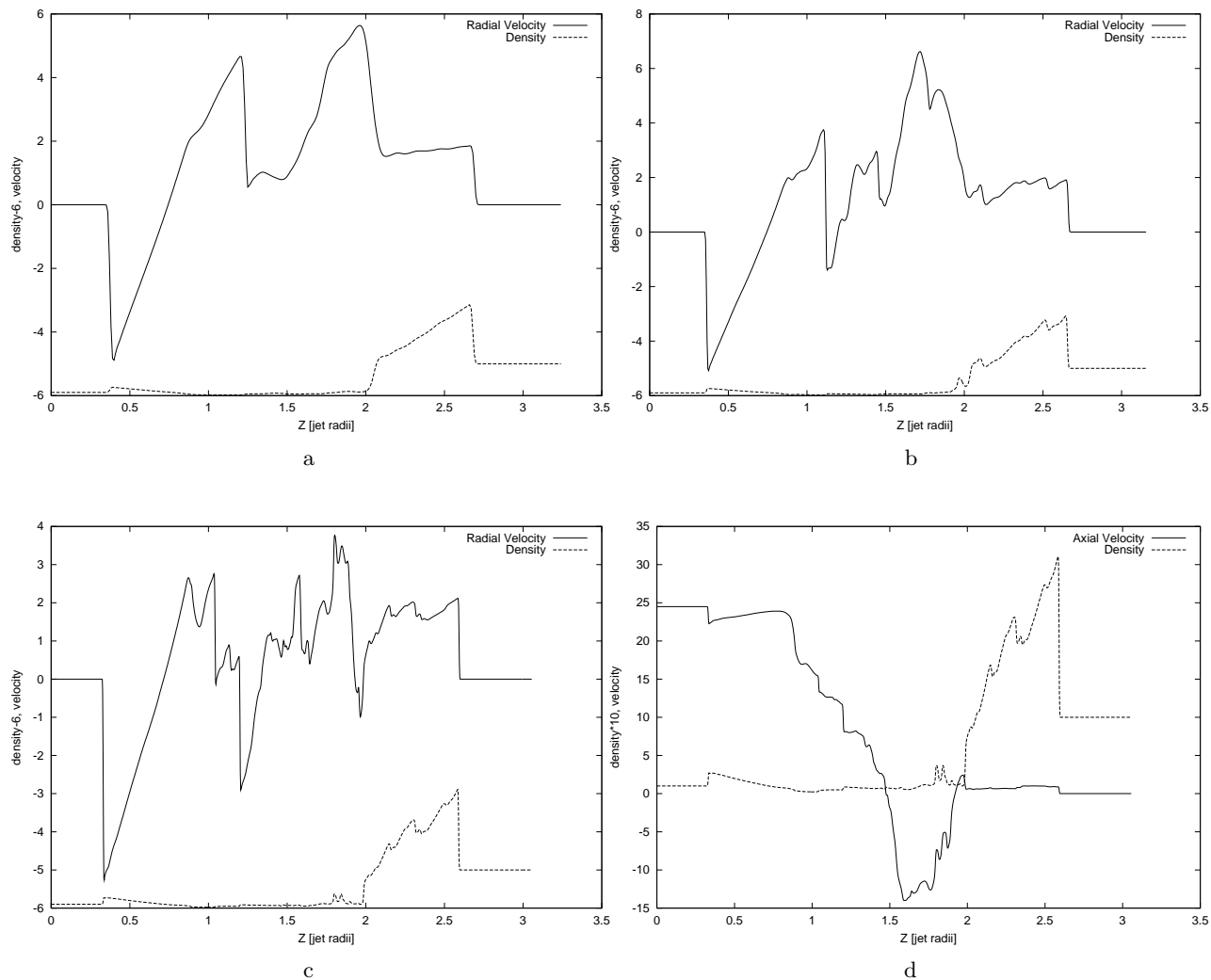

\begin{minipage}{17cm}
\rotatebox{-90}{\includegraphics[width=6cm]{fig09a.epsi}}%
\rotatebox{-90}{\includegraphics[width=6cm]{fig09b.epsi}}\\[0.2cm]
\mbox{}\hfill a \hfill\hfill b \hfill\mbox{}\\[0.4cm]
\rotatebox{-90}{\includegraphics[width=6cm]{fig09c.epsi}}%
\rotatebox{-90}{\includegraphics[width=6cm]{fig09d.epsi}}\\[0.2cm]
\mbox{}\hfill c \hfill\hfill d \hfill\mbox{}\\[0.4cm]
\end{minipage}
\caption{{\bf a-c:} Slices at $Z=1.5$ for radial velocity (solid)
   and density ($-6$) (dashed) over grid points at time t=0.5 for setup B. 
   The resolution is 100, 200, and 400 ppb for plot a to c, respectively.
   The shocked external medium is located between $R=2$ and $R=2.7$.
   {\bf d:} Radial slice for density ($\cdot 10$) 
   and axial velocity at t=0.5, Z=1.5 for the
   400 ppb simulation. At R=1.9 the fingers of the 
   KH instabilities are visible (increase in the density).
   They also appear in the velocity as slower regions.}
  \label{vxydslices}
\end{figure*}

At our highest resolution of 400 ppb we discover for the first time 
a completely different behavior of the KH instability.
A close look at Fig.~\ref{kmrescomp} shows that the amplitude of the 
dominant KH mode in the cocoon, which is continously
growing at lower resolution, is only about half as big 
in the 400 ppb plot as in the 200 ppb plot.
Furthermore, the contact discontinuity is turbulent almost everywhere,
already at earlier times, when the dominant mode was not yet evolved
(see lower left picture in Fig.~\ref{kmrescomp}).
This is quite contrary to the situation at lower resolution, where
the longest wavelength mode develops first. 
A gray scale plot of the logarithmic density for the three highest resolution
simulations at time $t \approx 0.5$
is shown in Figs.~\ref{kmrcgray_a}~and~\ref{kmrcgray_b}.
It shows that at 100 ppb the contact discontinuity between shocked 
ambient medium and jet backflow is smooth. This changes at 200 ppb.
Here at about $R=2$, evolved KH instabilities show
up over about one jet radius. In the magnification, one can see that a high 
density flow emerging from the triple shock point at the Mach disk,
which is also present at lower resolution, develops also 
KH instabilities. These two seem to interact with 
each other. In the 400 ppb gray scale plots, the bumps of the first one 
are located in immediate vicinity of the bumps of the other one.
In the 200 ppb plot at $R \approx 0.8$, the bump in the inner flow
even seems to hit the contact discontinuity. Furthermore, the two phenomena 
arise at the same resolution threshold.
The bumps in the contact discontinuity are not stationary. They move
to the left (see Fig.~\ref{vxydslices}d). 
The backflow of the jet gas in the vicinity of the 
contact discontinuity has a velocity of about 12 towards the left.
It accelerates the almost only outward moving shocked external medium
to velocities in excess of 7, which is about $30\%$ of the jet velocity.
Because of that motion, the KH instabilities in the head
region of the jet are always small in this early phase.
These moving KH instabilities act like a piston on the 
shocked external medium. They drive weak waves, 3 of which are seen in the 
lower left plot of Fig.~\ref{kmrescomp}. Indeed, their appearance can be
best seen in the radial slices of the density and radial velocity
(Fig.~\ref{vxydslices}a-c). The resolution comparison shows at 100 ppb in the 
region between $R \approx 2$ and $R \approx 2.7$
a nearly linear density increase. At 200 and 
400 ppb the peaks of the discussed waves are clearly visible.
Thus, it seems that the stream from
the Mach disk triple shock point causes
the KH instabilities at the contact discontinuity, which in turn
drives weak waves into the shocked external medium. 
Small KH instabilities are also observed at the beam boundary,
at highest resolution (Fig.~\ref{kmrescomp}).
They move towards the jet head. For example, the biggest bump in the lower 
plots of Fig.~\ref{kmrescomp} is located at $Z \approx 2$ in the left hand
plot and at $Z \approx 3$ in the right hand plot.
It depends on their velocity if they can be 
dangerous for the beam stability before they reach the jet head.
\subsubsection{Rayleigh-Taylor instabilities at the jet head}
\begin{figure*}
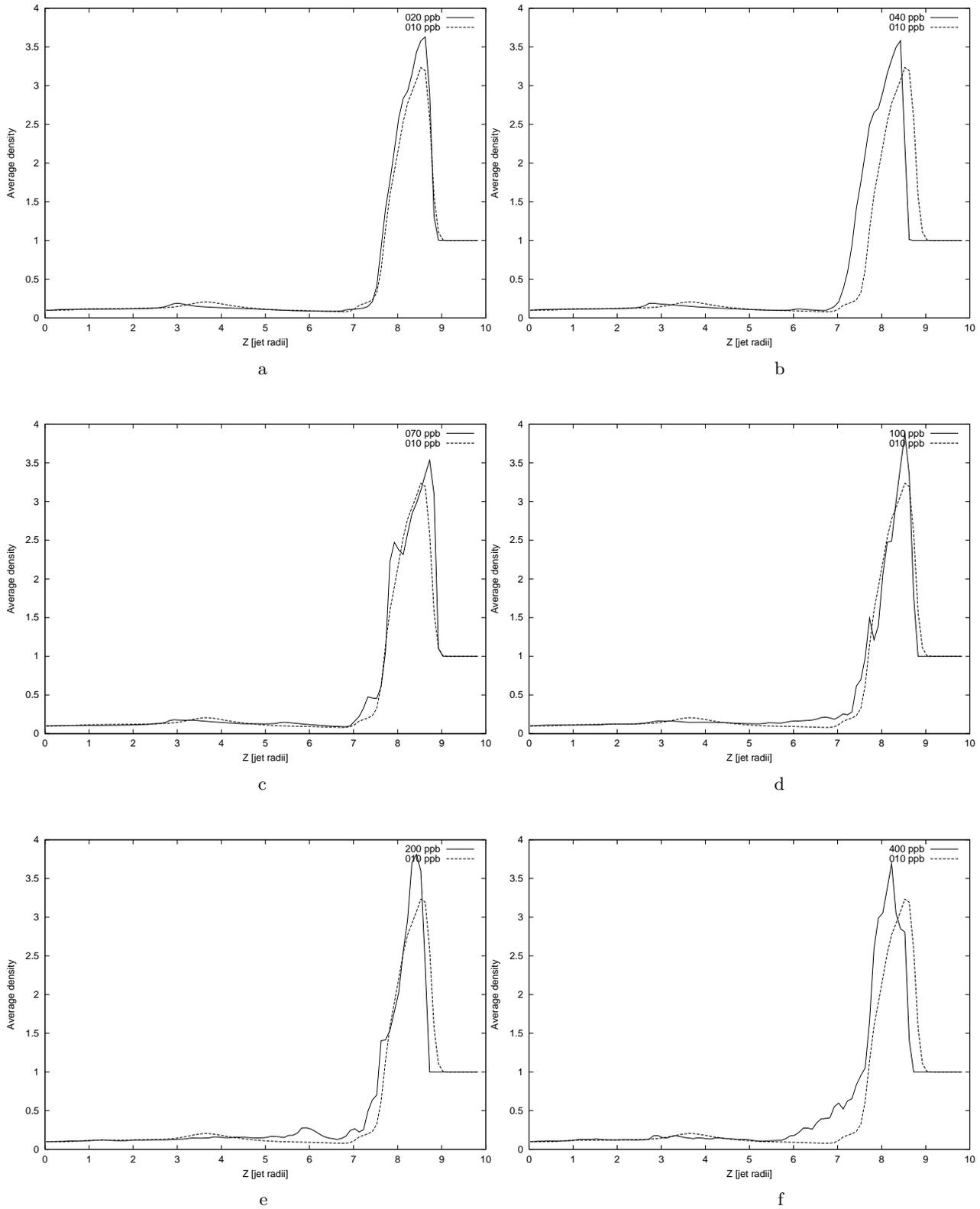

\rotatebox{-90}{\includegraphics[width=6cm]{fig16a.epsi}}%
\rotatebox{-90}{\includegraphics[width=6cm]{fig16b.epsi}}\\[0.2cm]
\mbox{}\hfill a \hfill\hfill b \hfill\mbox{}\\[0.4cm]
\rotatebox{-90}{\includegraphics[width=6cm]{fig16c.epsi}}%
\rotatebox{-90}{\includegraphics[width=6cm]{fig16d.epsi}}\\[0.2cm]
\mbox{}\hfill c \hfill\hfill d \hfill\mbox{}\\[0.4cm]
\rotatebox{-90}{\includegraphics[width=6cm]{fig16e.epsi}}%
\rotatebox{-90}{\includegraphics[width=6cm]{fig16f.epsi}}\\[0.2cm]
\mbox{}\hfill e \hfill\hfill f \hfill\mbox{}\\[0.4cm]
\caption{Density averaged over the beam  
radius for the simulations of setup B. Plots a to f apply
to 20 ppb, 40 ppb, 70 ppb, 100 ppb, 200 ppb and 400 ppb, respectively.
The mass entrainment into the jet head can be seen in plot f, between
$Z=6$ and $Z=7.5$. For comparison, the 10 ppb curve is plotted as a dashed 
line in each figure.}
\label{diffmass}
\end{figure*}

Furthermore, the ``beam pumping'' (see above) gives rise to the onset 
of Rayleigh-Taylor (RT) 
instabilities near the axis in the contact discontinuity,
which appear for the first time at a resolution
of 100 ppb in the original publication, but can be clearly 
identified in our 70 ppb
contour plot ($(r,z)\approx(0.5,7.5)$).
Close examination of the jet head (Fig.~\ref{kmrescomp}
,\ref{kmrcgray_a}, and \ref{kmrcgray_b})
reveals that also the development 
of RT instabilities is crucially dependent on the resolution.
The RT instability 
needs an accelerating jet, which only appears at a certain 
resolution threshold connected to the appearance of oblique shocks in the jet,
which are responsible for the acceleration. 
In order to investigate the mass entrainment into the jet head in more 
detail, we show the radial average of the density in the jet beam
(Fig.~\ref{diffmass}). In the low resolution plots, the contact discontinuity
is visible as a nearly vertical line joining the density in the beam
at nearly 90 degree. At 70 ppb small peaks appear at the contact 
discontinuity. This is the first sign from the RT instability.
At higher resolution the differences between five and eight jet radii 
become more pronounced. At 400 ppb, the density in the region between
$R=5$ and $R=7.5$ exceeds the density in the 10 ppb simulation by 0.2,
on average.
This corresponds to an entrained mass of about 
$\rho_\mathrm{m} R_\mathrm{j}^3$.  

\subsubsection{Convergence of global quantities}
\begin{figure}
  \resizebox{\hsize}{!}{\rotatebox{-90}{\includegraphics{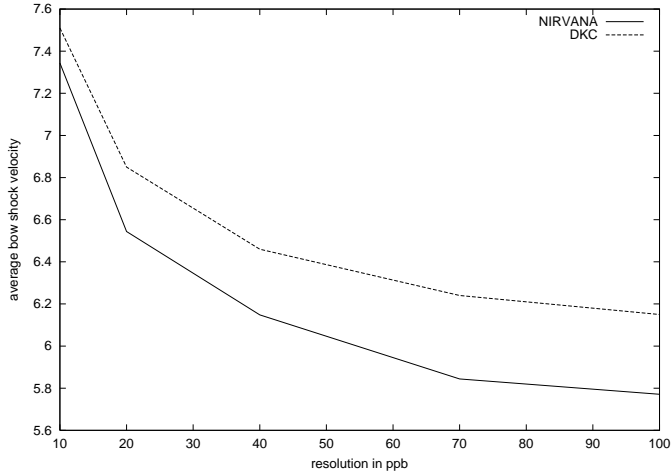}}}
  \caption{Comparison of the bow shock velocity of the setup B 
    jets with the ones 
    of equal snapshot time from \cite{Koessl1988}. The snapshot times
    were: 10 ppb: t=1.28, 20 ppb: t=1.41, 40 ppb: t=1.42, 70 ppb: t=1.99 and 
    100 ppb: t=1.98.}
  \label{kmvcmp}
\end{figure}

\begin{figure}
\resizebox{\hsize}{!}{\rotatebox{-90}{\includegraphics{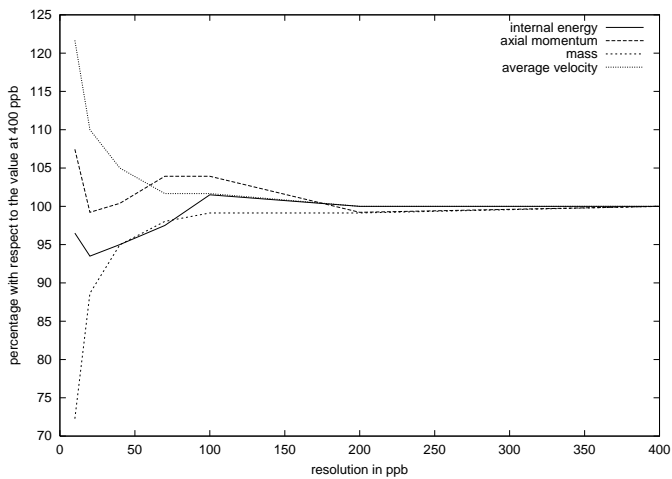}}}
  \caption{The total values of internal energy, axial momentum, mass and 
    average bow shock velocity against 
    resolution for the simulations from setup B. 
    At the resolution of 100 the mass is converged 
    with an accuracy of 
    3 digits. Here t=1.28 except for 70 ppb were t=1.27.}
  \label{kmrcglob}
\end{figure}

We have also computed some global quantities for each simulation.
Fig.~\ref{kmvcmp} shows that at every resolution 
our jet is slower than its counterpart
in the original publication by $\approx (4.5 \pm 2 )\%$. If we
use in NIRVANA a lower resolution by a factor of $2-2.5$ than the 
corresponding computation in the original publication we get the same 
average bow shock velocity. Fig.~\ref{kmrcglob} shows the convergence
of four global quantities up to the highest resolution. The first two are 
the bow shock velocity averaged over the computation time and the total mass
in the computational domain. These two parameters should be coupled because
the bow shock sweeps the mass, 
which is concentrated in the ambient medium, off 
the grid. They converge at 100 ppb. This reflects the fact that the
cocoon develops only additional small scale structure once 
the long wavelength instability has been sufficiently resolved.
The behavior changes at 400 ppb, where again more matter remains within the 
computational domain after a stagnation between 100 and 200 ppb. 
 Axial momentum is mainly situated in the region between
contact discontinuity and bow shock (shroud), the beam and
the backflow (decreasing order). In all the simulations compared in 
Fig.~\ref{kmrcglob}, the axial momentum changes by only 5\% indicating
that the global flow pattern is reproduced quite well already at low 
resolution. The internal energy is concentrated in the region 
behind the bow shock and in front of the Mach disk. This is why it is 
partly correlated to the axial momentum.
But convergence of this number indicates also a correct description
of the terminal shock region. As can be seen in Fig.~\ref{kmrescomp},
the terminal shock region seems to behave quite differently in simulations 
with different resolution. Changes of the internal energy 
on the 5\% level remain up to 200 ppb.  
The flat behavior between 200 and 400 ppb might be due to coincidence.

Summarizing, it seems that the simulation was well on its way to convergence
up to 200 ppb. But the surprising damping of the long wavelength 
KH instability opened up a new chapter in the convergence 
behavior.

\section{Magnetohydrodynamic jet simulations}
\subsection{Configuration}
The setup here is essentially the same as in setup C of the previous section
except for a toroidal magnetic field ($B_{\phi}$) and a jet pressure profile 
which assures initial transverse hydromagnetic equilibrium 
(see \cite{Lind1989} for details). The jet profile is:
\begin{equation}
B_{\phi}= \left\{ \begin{array}{r@{\quad,\quad}l}
      B_\mathrm{m} \; r/R_\mathrm{m} & 0 \le r < R_\mathrm{m} \\
      B_\mathrm{m} \; R_\mathrm{m}/r & R_\mathrm{m} \le r < R_\mathrm{j} \\
      0 & R_\mathrm{j} \le r
\end{array} \right.
\end{equation}
and
\begin{equation}
p= \left\{ \begin{array}{r@{\quad,\quad}l}
 \left[\alpha+\frac{2}{\beta_\mathrm{m}}\left(1-\frac{r^2}{R_\mathrm{m}^2}\right)\right] p_\mathrm{m} 
 & 0 \le r < R_m \\
 \alpha \;p_\mathrm{m} & R_\mathrm{m} \le r < R_\mathrm{j} \\
      p_\mathrm{m} & R_\mathrm{j} \le r
\end{array} \right.
\end{equation}
where $R_\mathrm{m}=0.37 (R_\mathrm{j})$, $B_\mathrm{m}=11.09$, $\alpha=0.33$, 
and $\beta_\mathrm{m}=0.205$.
The average plasma $\bar{\beta}$ which gives the ratio of the mean
internal gas pressure to mean internal magnetic pressure is defined as:
\begin{equation}
\bar{\beta}:=\frac{\bar{p}}{\frac{1}{8\pi}\frac{2}{R_\mathrm{j}^2}
\int_{0}^{R_\mathrm{j}}B^2_{\phi}(r^{\prime})\;r^{\prime}\;dr^{\prime}}
\end{equation}
and has the value of 0.6. 
The mean magneto-sonic Mach number, 
\begin{equation}
\bar{M}_\mathrm{j}=v_\mathrm{j} (\frac{2}{R^2_\mathrm{j}} 
\int_0^{R_\mathrm{j}} (\gamma+B_\mathrm{\phi}^2/(4 \pi p))\frac{p}{\rho}\;
 r\prime \; dr\prime)^{-1/2},
\end{equation}
is 3.5.
This corresponds to the highly magnetized jet model
in \cite{Lind1989}. 

\subsection{Comparison of results}
\begin{figure*}
 \centering
  \includegraphics[width=17cm]{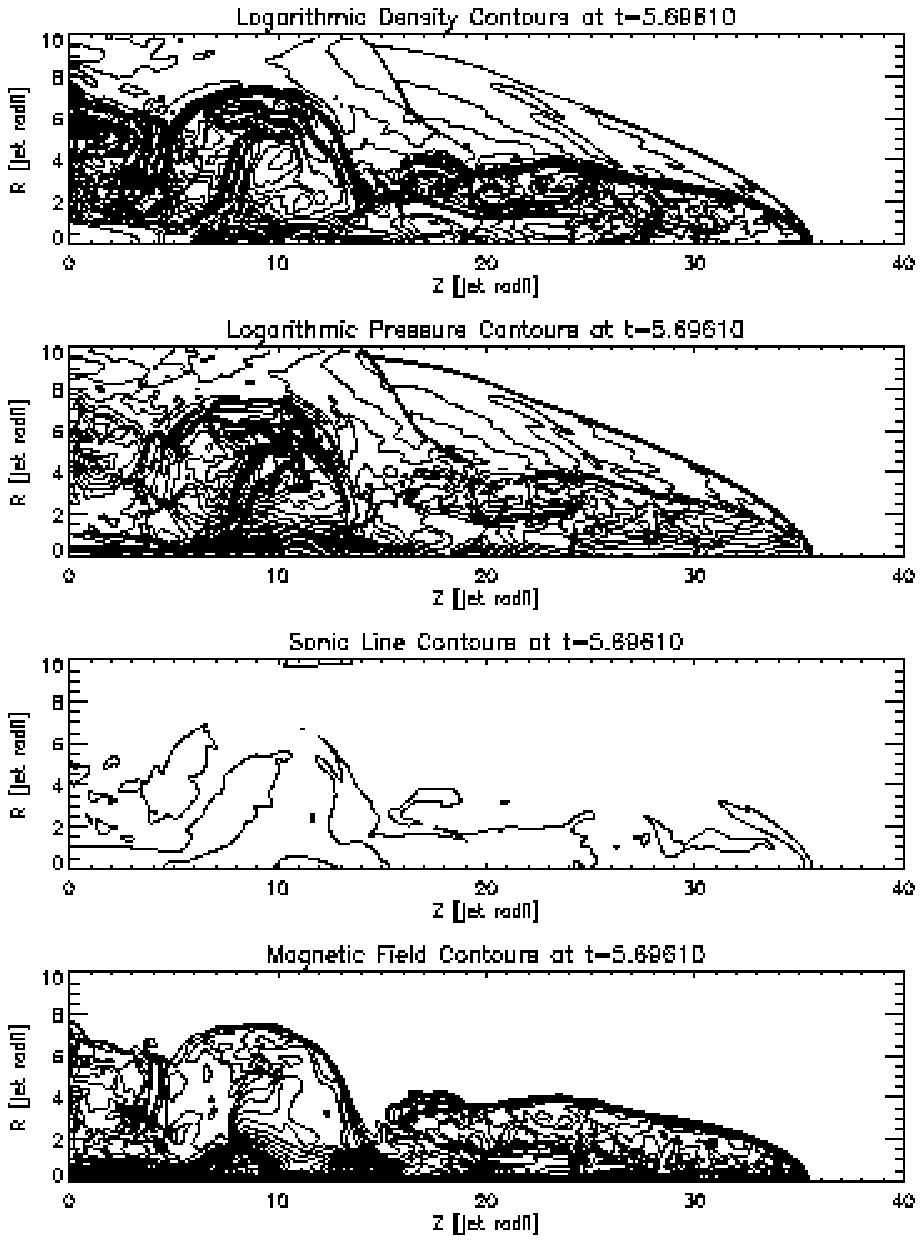}
  \caption{Contour plots of the density (30 logarithmically spaced lines),
    pressure (30 logarithmically spaced lines), magneto-sonic lines 
    (magneto-sonic Mach number equals 1 or -1) and toroidal magnetic field
    (30 lines) of the highly magnetized jet model at timestep 13500.
    This corresponds to a simulation time of 5.70. In units of 
    \cite{Lind1989} that would be 85.5.}
  \label{lchm19kc}
\end{figure*}
\begin{figure}
  \resizebox{\hsize}{!}{\rotatebox{-90}{\includegraphics{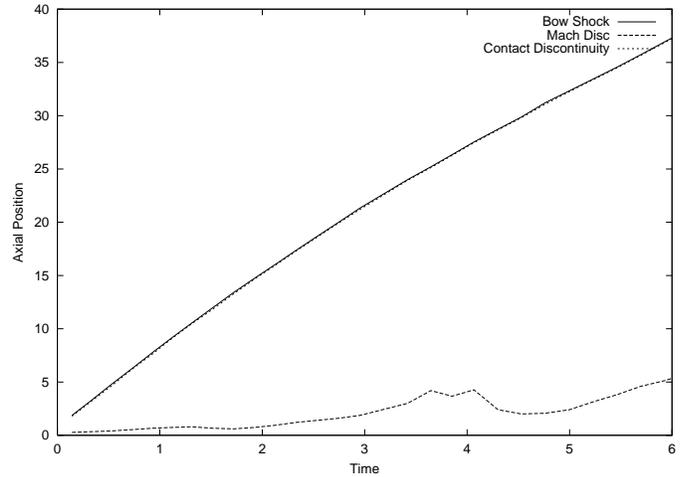}}}
  \caption{Time evolution of bow shock, Mach disk and contact discontinuity 
    for the simulation of Fig.~\ref{lchm19kc}. Notice that there is only a 
    small difference between the positions of contact discontinuity and bow 
    shock.}
  \label{lchmbcm}
\end{figure}

Detailed contour plots of the MHD jet at the end of the simulation are given 
in Fig.~\ref{lchm19kc}. At first sight, one recognizes a great similarity to
the corresponding picture in \cite{Lind1989}. The same nearly stationary 
big terminal vortex forms which includes the Mach disk. Here the jet material 
is driven away from the axis out to about $r=6$, where it is partly refocused 
onto the axis due to the Lorentz force, and another part manages to establish 
a backflow. This backflow turns again joining the main stream. 
As can be seen quite well in the sonic line plot the refocused 
material hits the z-axis at about $z=13$. There it separates itself again,
forming an on-axis backflow (which enhances the deflection from the axis
and then joins the main stream) and a time dependent outflow from
the region in the z-direction which leaves at $z \approx 15$. 
Due to the low pressure and the high magnetic field
($\beta \equiv 8 \pi p / {\bf B}^2 \approx 10^{-2}$), this area was called 
{\it magnetically dominated cavity} by \cite{Lind1989}. This is reproduced 
well in our simulation. When the plasma leaves this cavity, it forms a so 
called {\it nose-cone} of about $4 \,R_\mathrm{j}$ width, as it should be. This 
nose-cone ends at a contact discontinuity which can be seen in the plot of 
the toroidal magnetic field. At $t=5.7$ (Fig.~\ref{lchm19kc}) the bow shock 
has an average velocity of 6.27 which is about 3\% slower than the 
corresponding jet in \cite{Lind1989}. This might be due to the accuracy 
of the measurement which was carried out using an ordinary ruler for 
z-position of the bow shock in Lind's publication and dividing it by the 
simulation time. The advance of bow shock, Mach disk and contact discontinuity
is shown in Fig.~\ref{lchmbcm} and is generally very similar to 
the corresponding picture in \cite{Lind1989}. The exception 
is the position of the Mach disk, which has a considerably lower z-value
in our simulation. At $t\approx5.7$ our Mach disk has reached $z\approx5$
versus $z\approx7$ in the original publication. The advance of the Mach disk
seems to be coupled to the size of the on-axis backflow described above. 
\begin{figure*}
\centering
  \includegraphics[width=17cm]{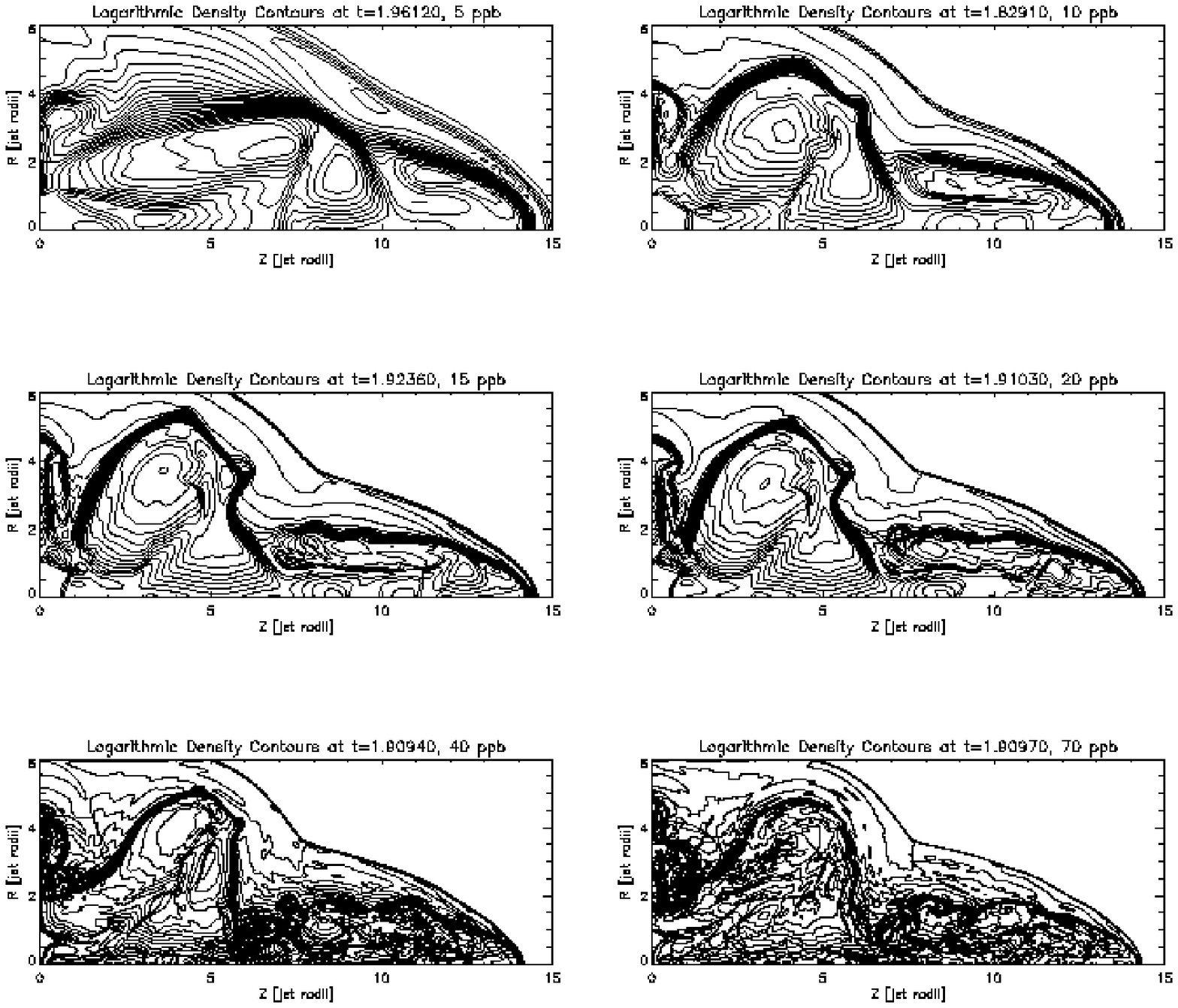}
  \caption{Snapshots of the logarithmic density contours of the MHD jet at 
    different resolutions. Times of the snapshots and resolution are indicated 
    on top of the individual figures.}
  \label{lchmrc}
\end{figure*}
\begin{figure}
  \resizebox{\hsize}{!}{\rotatebox{-90}{\includegraphics{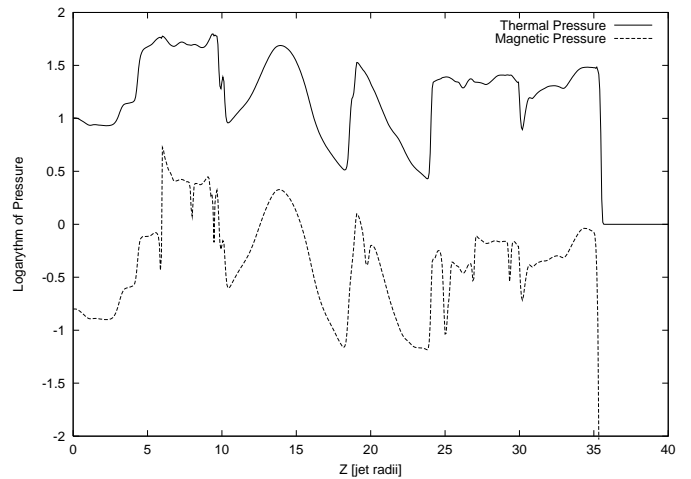}}}
  \caption{On axis ($r=1/15$) gas and magnetic pressure for the simulation
    of Fig.~\ref{lchm19kc}. The upper line displays the gas pressure and the 
    lower one the magnetic pressure. Note the close relationship between them.}
  \label{lchmpress}
\end{figure}
The amount of this on-axis backflow turns out to be sensible to the code used 
for the simulation: It is stronger in our simulation and therefore the Mach 
disk moves slower.
We have plotted gas and magnetic pressure close to the axis in 
Fig.~\ref{lchmpress}. One can see the close correlation of the magnetic pressure
and the gas pressure. Directly on the axis the magnetic field vanishes 
because of axi-symmetry. Therefore the magnetic field is 
generally weak in vicinity to the jet axis. The plot confirms also the 
original publication: Behind the Mach disk the gas pressure rises up to
60 and in the nose-cone it is on average $25 - 30$. 
In all our magnetized jet simulations, we do not find any RT 
instability. In contrast, we do find one prominent KH
instability developing at the upper right edge of the big vortex, probably 
excited by the jet stream that hits the contact discontinuity here. This was 
not observed in the original publication. The instability circled around the 
vortex and deposited an amount of jet plasma to the left of the vortex
where one can see shocked ambient gas in the original publication.
Other differences are the shock structures in the nose-cone. They are 
sharper in the original which one can trace back to the lower amount 
of diffusivity in shock regions by the FLOW code (compare also setup C
of our hydrodynamic section).

\subsection{Convergence}
 To investigate the convergence behavior we repeated the simulation at lower
and at higher resolutions (5 ppb, 10 ppb, 20 ppb, 40 ppb and 70 ppb). 
The results are shown in 
Fig.~\ref{lchmrc}. The time chosen for the snapshots was approximately 1.9. 
At that time the KH 
instability is excited as a bump at $(r,z)\approx(3.5,6)$. Interestingly, this 
bump disappears at a resolution of 70 ppb like in the original publication.
The behavior of the internal structure of the nose-cone 
does not seem to converge.
The Mach disk moves slower at higher resolution. It has advanced
$\approx 0.2 \,R_\mathrm{j}$ less at 20 ppb than at 15 ppb 
in Fig.~\ref{lchmrc}. 
At 70 ppb the Mach disk has reached the inflow boundary. 
Already at 40 ppb the shape of the contact discontinuity changes.
This is probably due to the vicinity of the Mach disc to the inflow
boundary.
The retreat of the Mach disk is surprising. Given higher efficiency
of FLOW, we should compare the 20 or 40 ppb simulation to the FLOW
result. Therefore the two simulations seriously disagree on the propagation of
the Mach disk, and it seems that NIRVANA approaches convergence in a different
way than FLOW, at least, if a dominant magnetic field is present.   
Nevertheless the overall shape of the bow shock and the contact surface 
remains essentially the same.
Also the average bow shock velocity stays remarkably constant,
7.6 at 15 ppb and 7.5 at the others: It seems to be converged.
With an eye on the lower resolution plots we could say that even 
10 ppb are sufficient to catch the correct behavior at the 
contact discontinuity and the bow shock.
But there is a sharp transition to lower resolution. This tells us 
that the essential features
dictating the shape of the bow shock and essentially also 
of the contact discontinuity 
are of the order $1/10$ of a jet radius.

\section{Discussion}

We have carried out simulations of magnetized and unmagnetized astrophysical 
jets in 2D. In the pure hydrodynamic simulations, 
we showed by detailed examination of a time series which was compared to the 
simulation by Koessl \& M\"uller (1988) and by a recomputation of the model of 
\cite{Lind1989} that in principle each of the evaluated codes is able to 
produce similar results. However they do not achieve this 
result with the same resolution: DKC needs more than twice the resolution to 
achieve similar results compared to NIRVANA. NIRVANA in turn needs somewhat 
more than twice the resolution in order to achieve the same results as FLOW.
Our results have revealed that depending on the exact method of shock 
handling the effective resolution of MHD codes 
-- measured through the convergence of global variables and inspection 
of characteristic features in the contour plots by eye --
differs considerably
more than 
in the test calculations by Woodward and Colella (1984)
-- when applied to the jet propagation problem.
If one looks at results produced by the codes at moderate resolution
(20 - 40 ppb) we find a characteristic representation of KH
instabilities: they appear as breaks in the contact discontinuity in DKC,
as round structures in FLOW and as intermediate a little bit unregular
but still round structures in NIRVANA.
While FLOW is quite an unusual code, because it needs no artificial 
viscosity,
it turns out to be the most efficient code by far, at least, if no
magnetic field is present. 
Strictly speaking, we cannot reproduce the results of FLOW with NIRVANA.
But the differences are explained by effects of resolution and artificial 
viscosity together.
We also have shown that the resolution of the simulation influences 
the average bow shock velocity  more in the non-laminar flow phase than in the 
laminar one. Because we explain the differences between the codes with a 
resolution effect, we conclude that in the laminar phase the beam structure 
is indeed converged whereas in the non-laminar one it is not. 
We find that the global jet parameters are converged in HD simulations 
with NIRVANA at $\approx$ 100 ppb. But even with our 
highest resolution computation (400 ppb) we do not 
achieve a fully converged beam structure: There are KH
instable regions in the cocoon and a complicated terminal shock structure
that seems to evolve in a turbulent manner. 
On the contrary, on the highest level of resolution qualitatively new 
behavior arises: the long wavelength KH instabilities 
are damped by the onset of small-scale turbulence,
which develops prior to the long wavelength modes, and the RT 
instability manages to entrain more and more mass into the jet's head with 
increasing resolution.
In a real situation however, one can expect that KH 
instabilities at the contact discontinuity will arise because of 
inhomogeneities in the external medium or a not completely steady jet flow.
The present computation shows that they arise, even with a homogeneous
external medium and a steady jet flow. 

Concerning the jet with a toroidal magnetic field we found 
a good convergence behavior up to 20 ppb: Already at 10 ppb the shape
of the bow shock and the contact discontinuity is essentially converged.
The average bow shock velocity changes only slightly up to 20 ppb.
(It remains constant for higher resolution.) 
A big problem is the discovered moving of the Mach disk towards the inflow 
boundary. It tells us that this particular simulation does not converge,
when computed with NIRVANA. It would be interesting to check 
this result for different initial conditions. 
A better configuration
for the jet with a toroidal magnetic field would probably be 
one with a time dependent injection like the outflow from the magnetically 
dominated cavity.
Furthermore this is a serious disagreement between FLOW and NIRVANA.
One possible explanation 
could be that the efficiency derived in the  hydrodynamic 
case can not be applied for simulations with a magnetic field,
maybe, because the different diffusivity description puts NIRVANA and FLOW
on two different convergence branches.
How the propagation of the Mach disk would change with resolution
for the FLOW case remains unclear.
Resolution studies with different codes are therefore desirable.

\begin{acknowledgements}
We acknowledge very helpfull comments by the referee.
This work was supported by the 
Deutsche Forschungsgemeinschaft
(Sonderforschungsbereich 437).
\end{acknowledgements}

{}
\end{document}